\documentclass[10pt,onecolumn,notitlepage,pra,groupedaddress]{revtex4-1}
\usepackage{graphicx}
\usepackage{epstopdf}
\usepackage{dcolumn}
\usepackage{bm}
\usepackage{amsmath}
\usepackage{amssymb}
\usepackage{mathrsfs}
\usepackage{amsfonts}
\usepackage{color}
\usepackage{dsfont}

\newcommand{\ket}[1]{\ensuremath{|#1\rangle}}
\newcommand{\bra}[1]{\ensuremath{\langle #1|}}
\newcommand{\braket}[1]{\ensuremath{\langle #1 \rangle}}
\newcommand{\op}[1]{\ensuremath{\hat #1}}
\newcommand{\opdag}[1]{\ensuremath{\hat{#1}^\dagger}}

\begin{document}

\title{Fully coupled hybrid cavity optomechanics: quantum interferences and correlations} 

\author{Juan Restrepo$^1$, Ivan Favero$^1$, and Cristiano Ciuti$^1$}
\address{$^1$ Laboratoire Mat\'eriaux et Ph\'enom\`enes Quantiques, Universit\'e Paris Diderot, CNRS UMR 7162, Sorbonne Paris Cit\'e, 10 rue Alice Domon et Leonie Duquet 75013 Paris, France}


\begin{abstract}
We present a quantum theory for a fully coupled hybrid optomechanical system where all mutual couplings between a two-level atom, a confined photon mode
and a mechanical oscillator mode are considered. In such a configuration, new quantum interference effects and correlations arise due to the interplay and competition between the different
physical interactions. We present an analytical diagonalization of the related fully coupled Hamiltonian, showing the
nature and energy spectra of the tripartite dressed excitations. We determine the driven-dissipative dynamics of such
hybrid systems and study phonon blockade effects under resonant excitation. We also study the statistical properties of the photon emission obtained under incoherent pumping of the two-level atom, which is particularly relevant for experiments with solid-state two-level emitters.

\end{abstract}


\maketitle

\section{Introduction}

Both cavity \cite{Haroche2006} and circuit \cite{Wallraff2004}  Quantum ElectroDynamics (cQED) have explored resonant light matter coupling between a confined photon mode and atom-like electronic excitations. On the other hand, quantum optomechanics \cite{Aspelmeyer2014, Favero2009} has explored off-resonant coupling between high-frequency photons and the motion of low frequency mechanical resonators. Superconducting circuits have demonstrated as well the possibility to strongly couple non-linear qubits directly to mechanical degrees of freedom \cite{laHaye2009, oConnell2010}. These fields have reached a point in their maturity allowing the study of hybrid structures presenting both cQED and optomechanical couplings, in both semiconductor structures \cite{Ding2010, Kimura2010, Jusserand2015} and superconducting circuits \cite{Lecocq2015,Pirkkalainen2015}, paving the way to multi-faceted optomechanical interactions.

From a theoretical point of view there have been many proposals exploring the physics yielded by these configurations. Hybridizing optomechanical systems and cQED with a single atom can lead to strong linear coupling between the motion of the atom and the mechanical resonator \cite{Hammerer2009}. Optomechanical cavities where the photon mode is coupled to an atomic ensemble have been studied \cite{Genes2008, Genes2011, Dantan2014, Carmele2014, Nie2016} and should lead to new properties of entanglement and cooling of mechanical motion. The coupling of the light field of an optomechanical cavity to a collective excitation such as a Bose-Einstein condensate \cite{deChiara2011} or a quantum well excitonic transition \cite{Kyriienko2014} has also been studied. More recently, the possibility to reach strong optomechanical coupling has opened the way to studies exploring hybrid systems containing a single two-level atom \cite{Restrepo2014,Sete2014, Wang2014,  Timo2015, Cernotik2015}.  

So far the considered physical couplings in hybrid tripartite cavity optomechanics are mediated by one component:  either the cavity being coupled both to the artificial atom and the mechanics, or the mechanics being coupled to both the cavity mode and the matter-like excitation. However, in solid state systems all three components can directly interact, calling for a more general description. In this paper we explore the physics of a fully coupled tripartite system, as depicted in Fig.1. In contrast to previous works, the studied situation accounts for all possible interference paths between atomic, photonic and mechanical states. As a consequence, new scenarios arise where dual couplings in the tripartite system can be effectively enhanced or canceled, producing unconventional non-linearities and photon-phonon correlations.

 The paper is organized as follows. In section \ref{sec:theModel} we introduce the model of the fully coupled hybrid cavity optomechanical system, consisting of an artificial two-level atom, a cavity mode and a mechanical resonator with all possible interactions within the tripartite system. In particular, in section \ref{analytical} we determine analytically the dressed excitations by diagonalizing the system Hamiltonian, while in section \ref{meq} we introduce the model of driving and dissipation, which is solved in the following sections. In section \ref{sec:effectiveCoupling} we discuss the interference effects occurring in the processes of cooling and heating, as a result of the interplay between the three different couplings within the tripartite system. In section \ref{sec:phononBlockade}, we present results of phonon blockade obtained under resonant coherent optical driving. In section \ref{sec:autoCorrelationFunctions}, we present our findings on spectral properties of the emission under incoherent pumping of the two-level system for such a tripartite system. Conclusions and perspectives are finally drawn in Sec. \ref{conclusions}.

\section{Fully coupled hybrid atom-cavity-mechanics system}
\label{sec:theModel}

\begin{figure}[h]
\centering
	\includegraphics[scale=1]{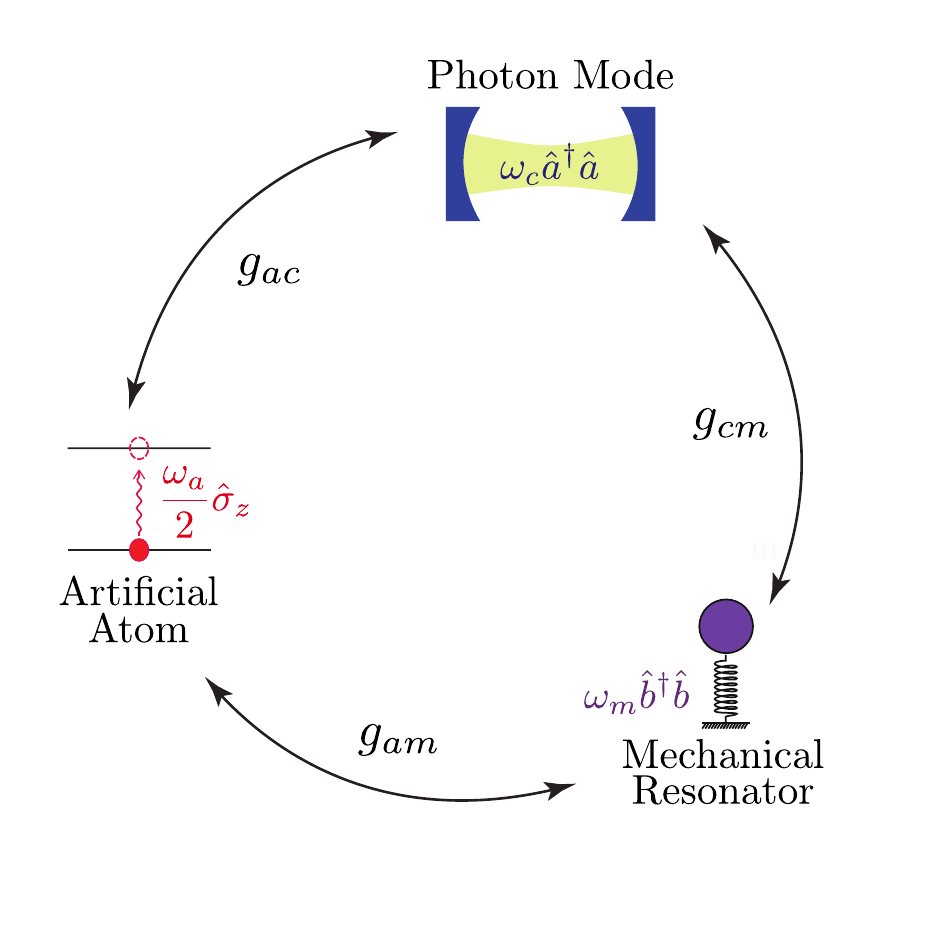}
	\caption{ (Color online) Scheme of the fully coupled tripartite hybrid system. A photon confined in a cavity of frequency  $\omega_c$, a two-level system with transition frequency $\omega_a$ and a mechanical resonator of frequency $\omega_m$ are coupled together. $g_{ac}$, $g_{cm}$ and $g_{am}$ represent the atom-cavity, cavity-mechanics and atom-mechanics coupling strengths respectively.}
	\label{fig:scheme}
\end{figure}


As can be seen in Fig.\ref{fig:scheme} we consider a hybrid system mixing usual cavity or circuit quantum electrodynamics (cQED) and optomechanics (OM). The system consists of an optical cavity mode, a two-level atom and a mechanical resonator. We model the cQED interaction in terms of a Jaynes-Cummings coupling between the cavity photons and the two-level system \cite{Haroche2006}. For the optomechanical part we consider the usual radiation-pressure coupling between the cavity mode and mechanical motion \cite{Aspelmeyer2014}. We also take into account a direct dispersive coupling \cite{WilsonRae2004} between the two-level system and the mechanical resonator. The corresponding Hamiltonian reads ($\hbar = 1$):

\begin{equation}
\begin{split}
	\op{H}_0 ~ =~&  \omega_c \opdag{a} \op{a} +\frac{\omega_a}{2} \op{\sigma_z} + \omega_m \opdag{b}\op{b}\\
		      & + i g_{ac} \left ( \op{\sigma}_+ \op{a} - \op{\sigma}_- \opdag{a} \right) \\
		      &  - g_{cm} \opdag{a}\op{a} \left ( \opdag{b} + \op{b} \right) - g_{am} \op{\sigma}_z \left ( \opdag{b} + \op{b} \right), 
\end{split}		   
\label{eq:Hamiltonian}
\end{equation}	

where $\op{\sigma}_{x,y,z}$ are the Pauli operators, $\op{\sigma}_-$ ($\op{\sigma}_+$) is the lowering (raising) operator for the two-level system, $\op{a}$ ( $\opdag{a}$ ) and $\op{b}$ ( $\opdag{b}$ ) are the annihilation (creation) operators of a photon and a mechanical excitation (phonon) respectively. The optical cavity and mechanical frequencies are $\omega_c$ and $\omega_m$ respectively. The frequency $\omega_a$ corresponds to the two-level transition. The Jaynes-Cummings interaction between the artificial atom and the cavity is quantified by the coupling term $g_{ac}$. $g_{cm}$ determines the cavity frequency shift produced by a displacement of the mechanical resonator in units of its zero-point motion amplitude. Finally $g_{am}$ accounts for the dispersive shift of the atom energy when the mechanical resonator is displaced.

\subsection{Analytical determination of the tripartite dressed excitations of the fully coupled system}
\label{analytical}
Here, we report on the analytical results showing the nature of the tripartite dressed excitations in the general fully coupled situation. We introduce the basis of eigenvectors of the uncoupled Hamiltonian ($g_{ac} = g_{cm} = g_{am} = 0$): $ \{  \ket{g, e} \otimes \ket{k} \otimes \ket{l} \}_{k,l \in \mathbb{N}}$. Here $\ket{g}$ ($\ket{e}$)  is the ket corresponding to the two-level system being in its ground (excited) state, $\ket{k}$ is the Fock state corresponding to $k$ photons in the cavity mode and $\ket{l}$ is the state with $l$ mechanical quanta. Let us introduce the operator counting the number of hybrid atom-cavity polaritonic excitations :

\begin{equation}
	\op{N}_{polariton} = \opdag{a}\op{a} + \op{\sigma_+} \op{\sigma_-}.
\label{eq:numberPolaritons}
\end{equation}	 

We note that this polariton number operator commutes with the total Hamiltonian of the system, $[ \op{H}_0, \op{N}_{polariton} ] = 0$. Hence, the Hamiltonian of the closed system is block-diagonal  in a basis of eigenvectors of the polariton number operator. We thus introduce the Jaynes-Cummings ladder of states diagonalizing the atom-cavity part of the Hamiltonian: $\op{H}_{JC} = \omega_c\opdag{a}\op{a} + \omega_a/2 \op{\sigma}_z + i g_{ac}(\op{\sigma}_+\op{a}-\op{\sigma}_-\opdag{a})$. Namely, we consider the following base for the atom-cavity subsystem $\{ \ket{G}, \ket{ \pm ^{(n)}} \}_{n \in \mathbb{N}^*}$, where

\begin{equation}
\begin{split}
	n= 0: & ~\ket{G}   = \ket{g} \otimes \ket{k=0}, \\
	\forall n \geq 1: &~ \ket{+^{(n)}}   = \cos\left(\frac{\phi^{(n)}}{2}\right) \ket{g, k=n} -i \sin\left(\frac{\phi^{(n)}}{2}\right) \ket{e, k=n-1}, \\
				&~ \ket{-^{(n)}}   = \sin\left(\frac{\phi^{(n)}}{2}\right) \ket{g, k=n} +i \cos\left(\frac{\phi^{(n)}}{2}\right) \ket{e, k=n-1},
\end{split}
\label{eq:polaritonStates}
\end{equation}

with $\phi^{(n)} \in ]-\pi/2, \pi/2]$ defined as $\tan(\phi^{(n)}) = 2 \sqrt{n}g_{ac}/(\omega_a-\omega_c) = 2\sqrt{n}g_{ac}/\Delta_{ac}$. The corresponding eigenenergies are given by:

\begin{equation}
\begin{split}
	\op{H}_{JC} \ket{G} &= -\frac{\omega_a}{2} \ket{G} = \omega_G \ket{G}, \\
	\op{H}_{JC} \ket{\pm^{(n)}} &= \left( \left( n-\frac{1}{2} \right) \omega_c \pm \sqrt{\frac{\Delta_{ac}^2}{4}+n g_{ac}^2} \right) \ket{\pm^{(n)}}\\
						& = \omega_{\pm}^{(n)} \ket{\pm^{(n)}}.
\label{eq:polaritonFrequencies}
\end{split}
\end{equation}	

For the sake of clarity in the notations we shift the energy origin so that in the remainder of this paper we have  $\omega_G = 0$. This induces the appearance of an additional $\omega_a/2$ energy in the Hamiltonian with respect to \cite{Restrepo2014}\\

Moving to the polariton-phonon basis $\{ \{\ket{G}, \ket{\pm^{(n)}} \}_{n,\in \mathbb{N}^*} \otimes \ket{l}\}_{l \in \mathbb{N}}$, the Hamiltonian becomes block diagonal and can be written in the following form:

\begin{equation}
\begin{split}
	 \op{H}_0 =  & \,\op{P}_G \Bigg ( \omega_m \opdag{b}\op{b} + g_{am} (\opdag{b} + \op{b})  \Bigg ) \op{P}_G \\
	 		    & + \sum_{n \geq 1}  \op{P}_n   \Bigg (  (n-1/2)\omega_c + \omega_a/2 + \sqrt{\frac{\Delta_{ac}^2}{4}+n g_{ac}^2} ~  \op{\sigma}^{(n)}_z  + \omega_m \opdag{b}\op{b}   - g_{cm}  (n-\frac{1}{2} ) (\opdag{b} + \op{b}) \\
			    & - \left(\frac{1}{2} g_{cm}- g_{am}\right) \left( \cos(\phi^{(n)}) \op{\sigma}_z^{(n)} + \sin(\phi^{(n)}) \op{\sigma}_x^{(n)}   \right)(\opdag{b} + \op{b})  \Bigg ) \op{P}_n .
\label{eq:projectedHamiltonian}
\end{split}	 
\end{equation}		

Here we have introduced $\op{P}_G = \ket{G}\bra{G}$ and $\forall n \geq 1, ~ \op{P}_n = \ket{+^{(n)}}\bra{+^{(n)}} + \ket{-^{(n)}}\bra{-^{(n)}}$, namely the projectors onto each subspace $\mathcal{H}_n$ containing states with exactly $n$ polaritons.  For $n \geq 1$ we also introduce Pauli matrices $\op{\sigma}_{x,y,z}^{(n)}$ acting on the 2-dimensional Hilbert space spanned by the $n$-polariton doublet $\{ \ket{-^{(n)}}, \ket{+^{(n)}} \}$. 

Once in the polariton-phonon basis the cavity-mechanics and atom-mechanics couplings lead to two distinct processes in each subspace $\mathcal{H}_n$. On the one hand we have a static shift of the mechanical resonator position (last term in the second line in Eq.\ref{eq:projectedHamiltonian}). The states spanned by the $n$-polariton doublet contain in average $n-1/2$ photons that displace the position of the mechanical resonator via radiation pressure. On the other hand, an additional exchange between the atom and the cavity mode leads to an effective coupling between the polariton states and the mechanical resonator of the form $\op{\sigma}^{(n)}(\opdag{b}+\op{b})$. For large atom-cavity detuning ($\Delta_{ac} \gg g_{ac}$ and hence $\phi^{(n)} \sim 0$) this results in a dispersive coupling between the mechanical resonator and the hybrid atom-cavity polariton excitations $(\propto \op{\sigma}_z^{(n)}(\opdag{b}+\op{b}))$. In contrast, for small detunings ($\Delta_{ac}\ll g_{ac}$) the hybridization between atomic and photonic excitations is more important and the effective coupling between the mechanical degree of freedom and the $n$-polariton doublets is then Rabi-like $(\propto \op{\sigma}_x^{(n)}(\opdag{b}+\op{b}))$. In the remainder of this article, we assume a resonant Jaynes-Cummings configuration ($\omega_c=\omega_a$) for which the Hamiltonian in the $n$-polariton subspace is simplified as:

\begin{equation}
	\op{H}^{(n)} = n \omega_c + \sqrt{n}g_{ac} \op{\sigma}_z^{(n)} + \omega_m \opdag{b}\op{b} - g_{cm}(n-1/2) (\opdag{b}+\op{b}) -\left(\frac{1}{2}g_{cm}-g_{am} \right) \op{\sigma}_x^{(n)}(\opdag{b}+\op{b})
\label{eq:nPolaritonHamiltonian}
\end{equation}

From Eq.\ref{eq:nPolaritonHamiltonian} it is possible to gain further insight into the eigenvectors and energies of the problem. Given the block diagonal structure of the Hamiltonian it is possible to diagonalize it by considering each subspace $\mathcal{H}_n$ separately. Let us introduce the effective polariton-mechanics coupling $\tilde{g}_{pm} = g_{cm}-2g_{am}$. In order to tackle the static shift of the mechanical resonator equilibrium position we start by defining displaced creation and annihilation operators for the mechanical resonator:

\begin{equation}
	\opdag{b}_n = \opdag{b}-\frac{q_0^{(n)}}{2}, 	\op{b}_n = \op{b}-\frac{q_0^{(n)}}{2},
\label{eq:displacedOperators}
\end{equation}	

where $q_0^{(n)} = (n-1/2) \sqrt{2} g_{cm}/\omega_m$  is the new equilibrium position for the states in $\mathcal{H}_n$. We associate to these operators a displaced basis of Fock states: $\{\ket{m^{(n)}}\}_{m \in \mathbb{N}}$, $\ket{m^{(n)}}$ being  the Fock state with $m$ phonons for a mechanical resonator centered at the position $q_0^{(n)}$. Rewriting the Hamiltonian of Eq.\ref{eq:nPolaritonHamiltonian} in this displaced frame leads to:

\begin{equation}
\begin{split}
		\op{H}^{(n)} = & ~\omega_m \opdag{b}_n\op{b}_n + \sqrt{n} g_{ac} \op{\sigma}_z^{(n)} - \frac{\tilde{g}_{pm}}{2} \op{\sigma}_x^{(n)}(\opdag{b}_n + \op{b}_n)\\
				      & -\omega_m \left( \frac{g_{cm}}{\omega_m} \right) ^2 (n - 1/2) \op{\sigma_x}^{(n)} \\
				      & + n \omega_c - \omega_m \left( \frac{g_{cm}}{\omega_m} \right) ^2 (n-1/2).
\end{split}
\label{eq:displacedHamiltonian}
\end{equation}			
	
In the displaced frame the static shift of the mechanical resonator leads to a perturbative effect $\propto \op{\sigma}_x^{(n)}$ inducing a Stark shift of the energies. This second order effect can be fully taken into account at the expense of losing analytical expressions for the eigenenergies and eigenvectors \cite{Braak2011} and will be neglected in the following analytical derivation. By performing a rotating wave approximation on the Rabi-like coupling between the $n$-polariton states and the displaced mechanical resonator we obtain a Jaynes-Cummings coupling between the two subsystems, which is diagonalized by the following basis of hybrid atom-cavity-mechanics states:

\begin{equation}
\begin{split}
	 \mbox{for }m = 0: & ~\ket{G_n}  = \ket{-^{(n)}}\ket{0^{(n)}}, \\
	\mbox{for }m \geq 1: &~ \ket{+_{n,m}}  =  \cos\left( \frac{\theta_{n,m}}{2} \right) \ket{+^{(n)}} \ket{(m-1)^{(n)}} + \sin\left( \frac{\theta_{n,m}}{2} \right) \ket{-^{(n)}} \ket{m^{(n)}}, \\
				&~ \ket{-_{n,m}}  =  \sin\left( \frac{\theta_{n,m}}{2} \right) \ket{+^{(n)}} \ket{(m-1)^{(n)}} - \cos\left( \frac{\theta_{n,m}}{2} \right) \ket{-^{(n)}} \ket{m^{(n)}},
\end{split}
\label{eq:polaronStates}
\end{equation}

where

\begin{equation}
	\tan[ \theta_{n, m}] = -\frac{\tilde{g}_{pm} \sqrt{m}}{2g_{ac} \sqrt{n}-\omega_m},~\theta_{n, m} \in \left]\frac{-\pi}{2},  \frac{\pi}{2}\right].\\
\end{equation}	

The corresponding energies for $n \geq 1$ and $m = 0$
\begin{equation}
	\bra{G_n} \op{H}^{(n)}\ket{G_n} = n \omega_c -\frac{g_{cm}^2}{\omega_m} \left( n-\frac{1}{2} \right) - g_{ac} \sqrt{n} ,
\label{eq:approxEigenEnergies_zeroM}
\end{equation}	
while for $n \geq 1$ and $m \geq 1$ we have
\begin{equation}
\label{eq:approxEigenEnergies_nonZeroM}
\begin{split}
	\bra{\pm_{n,m}}	 \op{H}^{(n)} \ket{\pm_{n,m}}  = & ~ n\omega_c -\frac{g_{cm}^2}{\omega_m} \left( n-\frac{1}{2} \right) + \left( m-\frac{1}{2} \right) \omega_m \\
												& \pm \sqrt{
												\frac{(2 g_{ac} \sqrt{n}-\omega_m)^2}{4} + m\frac{\tilde{g}_{pm}^2}{4}
												}\\
												= &\,  \omega_{\pm, n, m}, \end{split}
\end{equation}	

where $\omega_{\pm, n, m}$ is the Hamiltonian eigenvalue associated to the eigenvector $\ket{\pm_{n,m}}$.\\

 Whenever the $n$-polariton doublet is at resonance with the mechanical resonator $(2\sqrt{n}g_{ac} = \omega_m)$ the matrix elements linking the eigenstates in Eq.\ref{eq:polaronStates} to the original uncoupled basis are given by:

\begin{equation}
\begin{split}
	\braket{\pm_{n,m} | e,k,l} = & \frac{i}{2} \delta_{k, n-1} \left( \braket{m^{(n)} | l} \mp \braket{(m+1)^{(n)} | l} \right), \\
	\braket{\pm_{n,m} | g,k,l} = & \frac{1}{2} \delta_{k, n-1} \left( \braket{m^{(n)} | l} \mp \braket{(m+1)^{(n)} | l} \right).
\end{split}
\end{equation}	

The overlap between the Fock states of the original and displaced basis varies according to \cite{Smith2004} : 	

\begin{equation}
	\forall m,l \in \mathbb{N}, ~\langle l | m^{(n)} \rangle= \sqrt{\frac{m!}{l!} } \left( \frac{q_0^{(n)}}{\sqrt{2}} \right) ^{l-m} L_{m}^{l-m}\left( (q_0^{(n)}/\sqrt{2})^2 \right) e^{-(q_0^{(n)}/\sqrt{2})^2},
\label{eq:overlapDisplacedHarmonicOscillators}	
\end{equation}	

where $x \mapsto L_{m}^{l-m}(x)$ is the generalized Laguerre polynomial of degree $m$ and index $l-m$.\\

\begin{figure}[h]
\centering
	\includegraphics[scale=1.00]{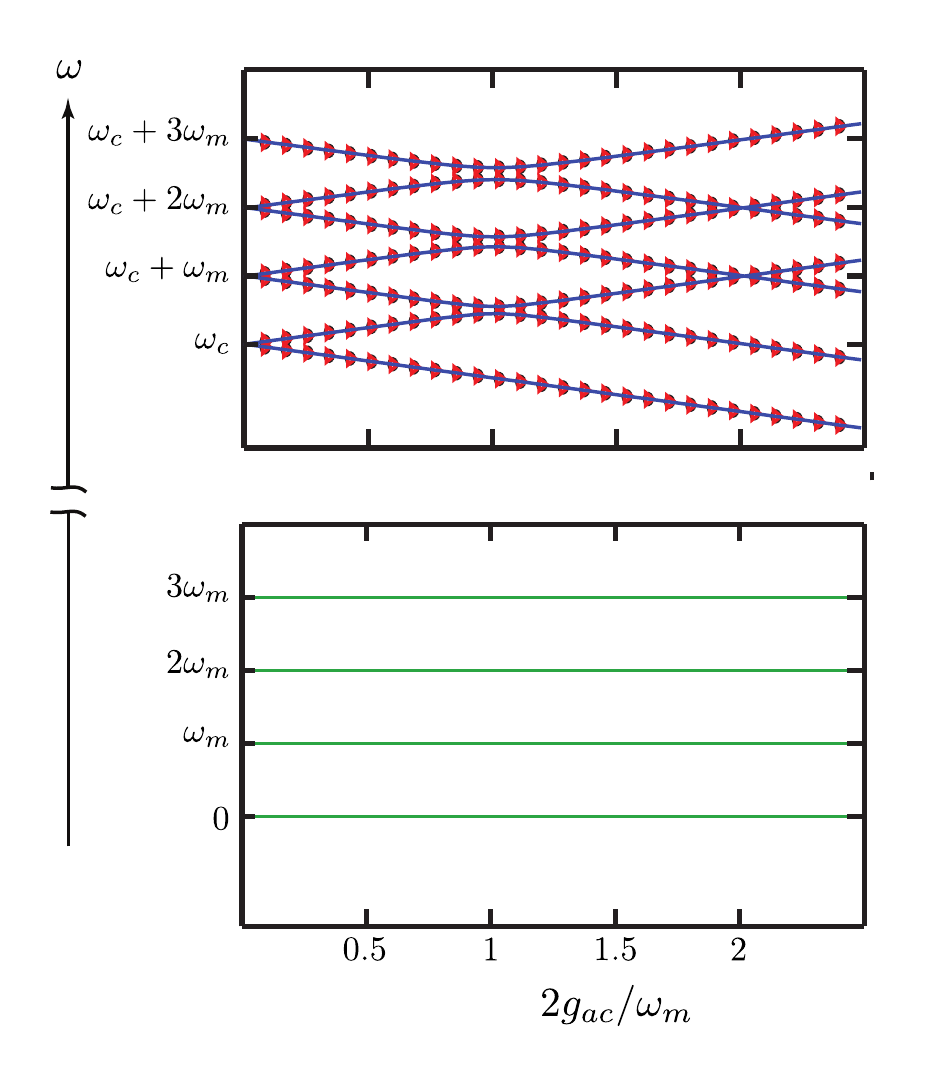}
	\caption{ (Color online) Energy levels of the Hamiltonian $\op{H}_0$ for polariton excitation numbers $n=0$ (bottom panel) and $n=1$ (top panel) as a function of the atom-cavity coupling $g_{ac}$ at resonance ($\omega_c=\omega_a$).  For the $1$-polariton subspace, blue solid lines have been calculated via the analytical expressions in Eqs.\ref{eq:approxEigenEnergies_zeroM} and \ref{eq:approxEigenEnergies_nonZeroM} (only the first 7 levels are plotted here). Black circles and red triangles have been obtained through a numerical diagonalization of the Hamiltonians in Eqs.\ref{eq:Hamiltonian} and \ref{eq:projectedHamiltonian} respectively. Other parameters: $g_{cm}/\omega_m = 1/20$ and $g_{am}/\omega_m = -1/40$,  giving an effective polariton-mechanics coupling $\tilde{g}_{pm}=\omega_m/10$.}
	\label{fig:energyLevels}
\end{figure}


Fig. \ref{fig:energyLevels} presents the eigenfrequencies of the Hamiltonian under consideration as a function of the atom-cavity coupling $g_{ac}$ for a resonant Jaynes-Cummings scenario $(\omega_a = \omega_c)$,  for 0 and 1 polaritons. The plot reports the results obtained by numerical diagonalization of the starting Hamiltonian in Eq. (1), the block-diagonal Hamiltonian in Eq. (5) and the  approximated analytical expressions in  Eqs.\ref{eq:approxEigenEnergies_zeroM} and \ref{eq:approxEigenEnergies_nonZeroM}. The agreement between the different calculations is excellent (the difference between the curves is not visible in the graph).
In the $0$-polariton subspace (lower panel of Fig.\ref{fig:energyLevels}) the mechanical resonator is decoupled from the cavity and the two-level atom. The corresponding energy levels follow the usual harmonic structure without any dependance on $g_{ac}$. The picture is radically different in the $1$-polariton subspace (upper panel of Fig.\ref{fig:energyLevels}). In this case the mechanical resonator and the non-linear $1$-polariton doublet are coupled. We obtain an anharmonic level structure arising from the energy splitting described in Eq.\ref{eq:approxEigenEnergies}. Changing the atom-cavity coupling results in a modification of the polariton energy splitting, which brings in and out of resonance the atom-cavity polaritonic transitions and mechanical vibrations. This ultimately results in a modulation of the hybrid atom-cavity-mechanics energy levels of the system leading to multiple crossings and anti-crossings.\\

\subsection{Theoretical framework in presence of driving and dissipation}
\label{meq}
In order to  discuss the dynamics of the open system in presence of  a coherent optical drive and of optical, mechanical and atomic losses, we consider the system density matrix $\op{\rho}$ and the following Lindblad master equation ruling its time evolution:

\begin{equation}
\begin{split}
	\frac{d \op{\rho}}{dt} = & -i \left[ \op{H}_0 + \op{V}_p(t) , \op{\rho} \right] \\
					& + \gamma_{c} L[\op{a}]\op{\rho} + \gamma_{a} L[\op{\sigma}_-] \op{\rho} \\
					& +n_{th} \gamma_m L[\opdag{b}] \op{\rho} + (n_{th} + 1) \gamma_m L[\op{b}] \op{\rho}.
\end{split}	
\label{eq:ME}
\end{equation}

Here $\gamma_m,~\gamma_c$ and $\gamma_a$ are the mechanical, cavity and atomic loss rates respectively. We suppose that the environment is in thermal equilibrium at temperature $T$. $n_{th}$ is the thermal mean phonon occupancy. We assume that the atom and cavity frequencies at play are such that from a polariton point of view the environment is at zero temperature $\omega_{a,c} \gg k_b T$, with $k_b$ Boltzmann's constant. $\op{V}_p(t) = i F_p (\opdag{a} e^{i\omega_p t} -\op{a} e^{-i\omega_p t})$ describes the coherent optical drive with frequency $\omega_p$ and amplitude $F_p$. We use the following convention for the Lindblad dissipative terms: $\forall \op{c}, L[\op{c}]\op{\rho} = \op{c} \op{\rho} \opdag{c} -1/2(\opdag{c}\op{c}\op{\rho}+\op{\rho}\opdag{c}\op{c})$. In the following we assume for simplicity that the atom and cavity loss rates are equal $\gamma_c=\gamma_a=\gamma_{ac}$ and we introduce the polariton and mechanical quality factors $Q_{ac}=\omega_{a,c}/\gamma_{ac}$ and $Q_m = \omega_m/\gamma_m$. An incoherent atomic excitation of the system will be considered in section \ref{sec:autoCorrelationFunctions}.\\

\section{Interference effects and correlations}
\label{sec:effectiveCoupling}

In Ref.\cite{Restrepo2014} we discussed how the insertion of a two-level system inside an optomechanical cavity modifies the usual optomechanical effects of  cooling and amplification for the mechanical motion. In that work the artificial atom was only coupled to the cavity ($g_{am}=0$). Here, we explore the physics of the tripartite hybrid system by including a direct coupling between the atom and the mechanics  ($g_{am} \neq 0$).\\

 In a standard atom-less optomechanical scenario, cooling (amplification) of mechanical motion is obtained by tuning the optical pump to the red (blue) optomechanical sideband of the cavity resonance \cite{WilsonRae2007, Marquardt2007}. In \cite{Restrepo2014} we showed that with an atom in an optomechanical cavity (with $g_{am}=0$) it is possible to favor similar processes by resonantly tuning the pump frequency in such a way to excite the Jaynes-Cummings states defined in Eq.\ref{eq:polaritonFrequencies}.  With the effective coupling between the polaritons and the mechanics discussed in Eq.\ref{eq:nPolaritonHamiltonian} these processes can be understood as exchanges of excitations between the polariton and mechanical subsystems. The Jaynes-Cummings subsystem is optically excited from its ground state  $\ket{G}$ to the lower (upper) polariton state. The polariton-mechanics coupling induces then an excitation (deexcitation) within the one-polariton doublet accompanied by absorption (emission) of a  phonon. The polaritonic excitation is then recycled via atom-cavity losses bringing the Jaynes-Cummings subsystem back to its ground state so that the cycle can start over. \\
 
 Following this picture, let us set $g_{am} \neq 0$ and go back to the effective polariton-mechanics coupling obtained in Eq.\ref{eq:nPolaritonHamiltonian}. From the expression of the coupling $\tilde{g}_{pm}$, matching the atom-mechanics coupling $g_{am}$ to half the value of the radiation pressure coupling $g_{cm}$ should cancel the effective polariton-mechanics interaction and thus hinder the effects on mechanical motion. 
\begin{figure}[h]
\centering
	\includegraphics[scale=1.00]{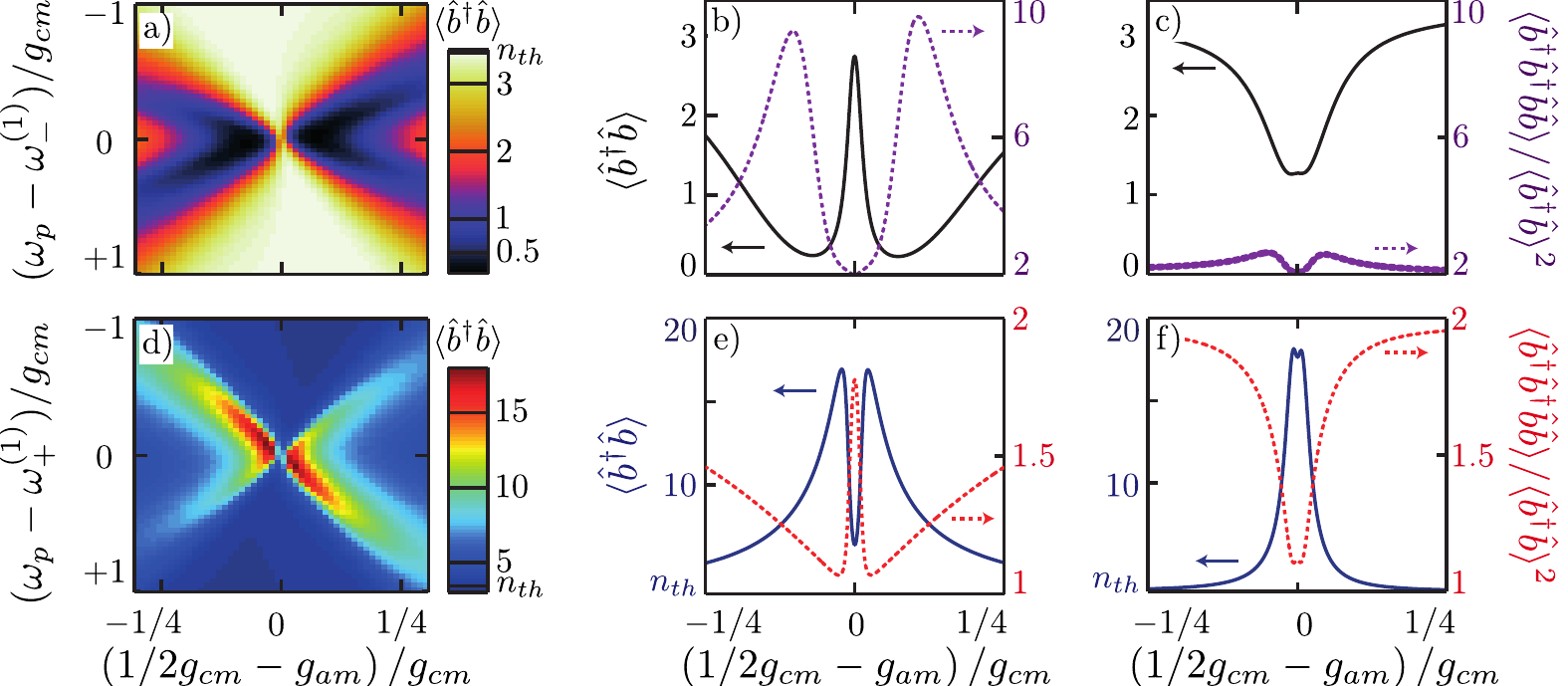}
	\caption{ (Color online) Optomechanical interferences and phonon correlations in a fully coupled hybrid tripartite system. {\bf a)} and {\bf d)} show the stationary number of mechanical excitations for pump frequencies around the lower $\omega_-^{(1)}$ (upper $\omega_+^{(1)}$) polariton frequency leading to cooling (amplification) of mechanical motion. The four panels on the right present the stationary number of phonons (solid lines) and mechanical second order autocorrelation function (dashed lines) as a function of the polariton-mechanics coupling, for different tones of the pump frequency $\omega_p$. {\bf b)} and {\bf c)}: Pump frequency set to excite the first and second order cooling transitions ($\omega_p \sim \omega_-^{(1)}$ and $\omega_p \sim \omega_-^{(1)} - \omega_m$) respectively. {\bf e)} and {\bf f)}: Pump frequency set to excite the first and second order amplification transitions ($\omega_p \sim \omega_+^{(1)}$ and $\omega_p \sim \omega_+^{(1)} + \omega_m$) respectively. In all six panels we consider the following set of parameters: $\gamma_{ac}/\omega_m=10^{-2}$, $g_{cm}/\omega_m=10^{-1}$, $Q_m = Q_{ac}=10^4$, $F_p/\gamma_{ac}=1$, $n_{th}=3.45$. }
	\label{fig:effectiveCoupling}
\end{figure}


Panels {\bf a)} and {\bf d)} of Fig.\ref{fig:effectiveCoupling} display the evolution of the stationary number of phonons (obtained by diagonalizing the Liouvillian super-operator associated to the master equation in Eq.\ref{eq:ME}) for a bath at finite temperature as a function of the pump frequency and the normalized polariton-mechanics coupling $(1/2 g_{cm}-g_{am})/g_{cm}$. It is worth noticing that at the exact value $g_{am}=1/2g_{cm}$ there is indeed a decrease in the efficiency of the cooling/amplification mechanisms.  This is clearly represented in panels {\bf b)} and {\bf e)} of Fig.\ref{fig:effectiveCoupling} where  $\omega_p = \omega_-^{(1)}$, $\omega_+^{(1)}$ respectively: in particular, the panels show the stationary number of phonons (solid lines) and mechanical second-order autocorrelation function (dashed lines).  

We previously discussed these processes in terms of an exchange of excitations between the $1$-polariton doublet and the mechanical resonator. A more accurate picture of the problem is drawn by looking closely at the transitions between eigenstates excited by the coherent pump. We consider the basis of eigenstates defined by Eq.\ref{eq:polaronStates}, which yields an effective theory valid for small values of $\tilde{g}_{pm}$. Let us suppose that the system is initially in a state of the form  $\ket{G}\ket{l}$ and that the pump is set to excite the lower polariton, $\omega_p \sim \omega_-^{(1)}$, (the case of $\omega_p \sim \omega_+^{(1)}$ will be discussed in section \ref{sec:phononBlockade}). The pump excites the system from subspace $\mathcal{H}_0$ to subspace $\mathcal{H}_1$ and the new system state is approximately $\ket{-^{(1)}} \ket{l}$, which is not an eigenstate of $\op{H}_0$. Due to the effective coupling between the polariton and the mechanics the two eigenstates actually excited by the pump are $\ket{\pm_{1,l}}$. The eigenenergies of this doublet are symmetrically positioned around $\omega_-^{(1)}$ , and separated by a frequency splitting $\sim \sqrt{l} \tilde{g}_{pm}$. When considering the open system, multiple transitions involving different values of $l$ have to be taken into account. As $\tilde{g}_{pm} \rightarrow 0$ the splitting between  levels decreases and a single tone of the pump frequency will be able to excite a larger number of transitions thus leading to the enhancement of cooling (or amplification). Nevertheless the exchange of an excitation between the two subsystems remains necessary, which is why we get an exact supression of these processes at $\tilde{g}_{pm}=0$.

In panels {\bf c)} and {\bf f)} of Fig.\ref{fig:effectiveCoupling} we tune the frequency to excite second-order processes ( in {\bf c)} $\omega_p=\omega_-^{(1)}-\omega_m$ for second-order cooling and in {\bf f)} $\omega_p=\omega_+^{(1)}+\omega_m$ for second-order amplification). Due to the non-resonant processes excited by such pump frequencies we observe a smaller contrast in the aforementioned effects.

\section{Phonon blockade under coherent optical drive}
\label{sec:phononBlockade}

Here we discuss more specifically  the case where the pump frequency is tuned in order to amplify mechanical motion ($\omega_p \sim \omega_+^{(1)}$). In particular we show that the coupling with the non-linear Jaynes-Cummings subsystem provides efficient anti-bunched phonon statistics under coherent optical drive, which is different from the the phonon dynamics  under incoherent excitation studied in Ref.\cite{Restrepo2014}.\\

\begin{figure}
\centering
	\includegraphics[scale=1.00]{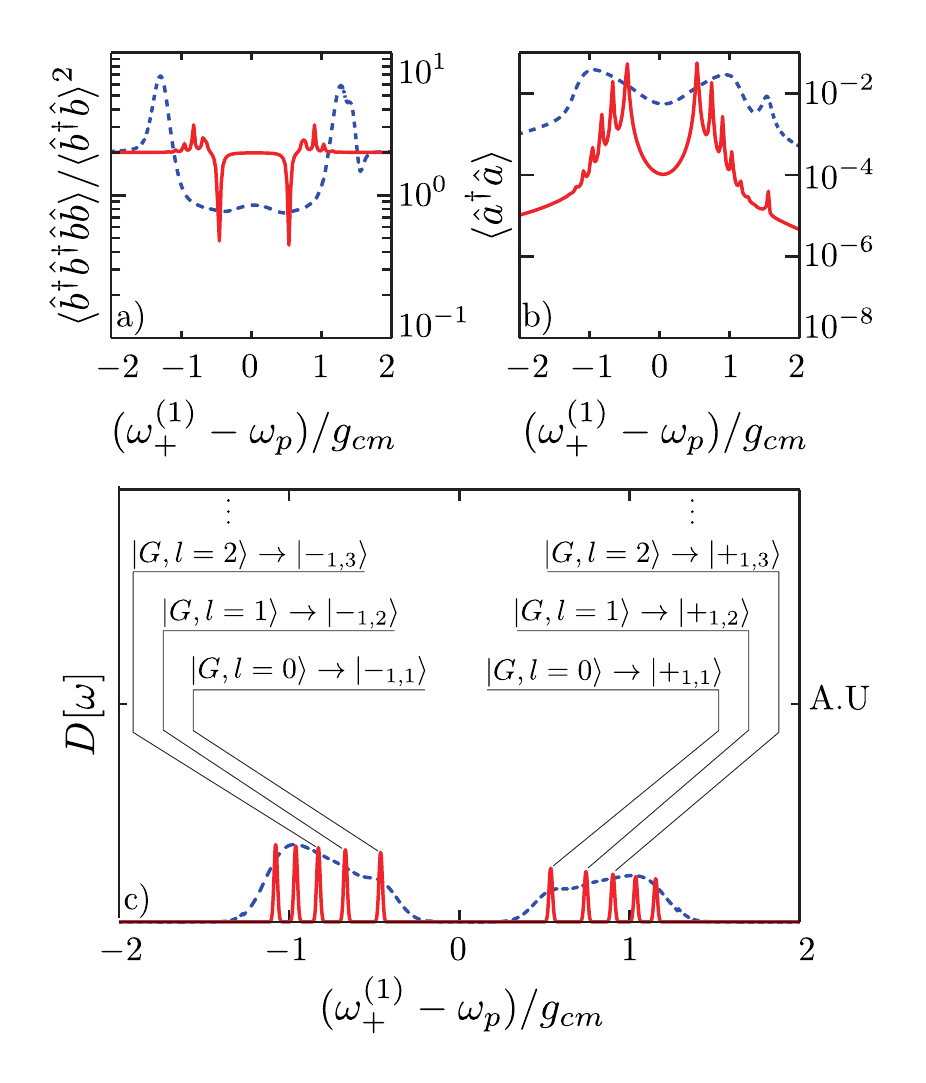}
	\caption{ (Color online) Coherently excited phonon blockade. a) Stationary phonon autocorrelation function, b) stationary number of photons and c) spectral joint density of states (convoluted by a Lorentzian of width $\gamma_c$) as a function of the pump frequency $\omega_p$ (in this plot, we only take into account transitions involving $m,l \leq 5$). All quantities are depicted around the upper 1-polariton frequency $\omega_+^{(1)}$. Red-solid and blue-dashed lines correspond to different values of the polariton quality factor $Q_{ac} = 10^5, 10^4$ respectively. In c) we present the transitions between eigenstates corresponding to each peak.  }
	\label{fig:phononBlockade}
\end{figure}

Panels {\bf a)} and {\bf b)} of Fig. \ref{fig:phononBlockade} present the mechanical second-order autocorrelation function $g^{(2)}(0) = \braket{\opdag{b}\opdag{b}\op{b}\op{b}}/\braket{\opdag{b}\op{b}}^2$ and stationary number of photons $\braket{\opdag{a}\op{a}}$ as a function of the pump frequency $\omega_p$ for values near the upper-polariton frequency. The environment temperature has been set to $T=0$. The red-solid line corresponds to the polariton quality factor $Q_{ac}=10^5$. The phonon autocorrelation function deviates from the value of a thermal distribution ($g^{(2)}(0)=2$) for specific values of the pump frequency. These correspond to the peaks on the stationary number of photons observed in panel {\bf b)}. We observe in particular that the first two peaks in the vicinity of the upper-polariton frequency lead to anti-bunching of phonons ($g^{(2)}(0) \sim 0.5$). For lower values of the polariton quality factor (blue-dashed line, $Q_{ac}=10^4$) the different peaks are no longer resolved and the autocorrelation functions tends to that of a coherent distribution ($g^{(2)}(0) \sim 1$).\\

To give further insight into these phenomena, let us assume  that the system is initially in a state of the form $\ket{G}\ket{l}$. Now the pump is tuned to the upper polariton energy, hence the system is promoted to $\ket{+^{(1)}}\ket{l}$: again, this is not an eigenstate of the Hamiltonian, but a linear superposition of the eigenstates $\ket{\pm_{1,l+1}}$. In order to fully understand these transitions we introduce the spectral joint density of states:

\begin{equation}
	D[\omega_p]=\sum_{\substack{s=\pm \\m, l \in \mathbb{N}}} |\bra{s_{1,m}}
	iF_p(\opdag{a}-\op{a})
	\ket{G,l}|^2 \delta [ \omega_p-(\omega_{s,1,m}-l\omega_m)  ].
	\label{eq:oSPJD}
\end{equation}

We plot this quantity in panel {\bf c)} of Fig. \ref{fig:phononBlockade}. For the sake of clarity we have convoluted the function $D[\omega_p]$ by a Lorentzian of width $\gamma_{ac} = \omega_{ac}/10^5$ (red-solid line) or $\gamma_{ac} = \omega_{ac}/10^4$ (blue-dashed line). We add labels indicating the transitions associated to each peak of the spectral joint density of states, for pump frequencies around the upper-polariton energy. It becomes clear that the anti-bunched statistics of phonons observed in panel {\bf a)} of Fig. \ref{fig:phononBlockade} arises from the excitation of states containing 1 hybrid atom-cavity-mechanics excitation $\ket{\pm_{1,1}}$. These states are a superposition of mechanical states with $0$ and $1$ phonon. On the other hand, as the polariton quality factor decreases (blue dashed lines), it is no longer possible to resolve each transition individually and the mechanical resonator ends in a more ``classical'' mixture of states leading to a coherent-like autocorrelation function ($g^{(2)}(0) \sim 1$).

\section{Optical emission: second-order auto-correlations of photons}
\label{sec:autoCorrelationFunctions}

Here we discuss statistical optical emission properties of the system under incoherent driving of the atom. Such a configuration is within reach for experimentalists dealing with solid-state implementations of the system under discussion (such as embedded semiconductor quantum dots, diamond  NV centers or superconducting circuit QED systems). To treat such effects, we will consider the predictions of the master equation accounting for the incoherent driving of the two-level atom, namely: 

\begin{equation}
\begin{split}
	\frac{d \op{\rho}}{dt} = & -i \left[ \op{H}_0 , \op{\rho} \right] \\
					& + \gamma_{c} L[\op{a}]\op{\rho} + \gamma_{a} L[\op{\sigma}_-] \op{\rho} + \gamma_{inc}L[\op{\sigma}_+]\op{\rho}\\
					& +n_{th} \gamma_m L[\opdag{b}] \op{\rho} + (n_{th} + 1) \gamma_m L[\op{b}] \op{\rho},
\end{split}	
\label{eq:incoherentME}
\end{equation}

where we have introduced the incoherent driving rate $\gamma_{inc}$. Here, we will focus
on the optical second-order autocorrelation function in the steady-state regime:

\begin{equation}
	g^{(2)}(\tau) = \frac{\braket{\opdag{a}(t)\opdag{a}(t+\tau)\op{a}(t+\tau)\op{a}(t)}}{\braket{\opdag{a}\op{a}}^2}.
\label{eq:timeSecondOrder}
\end{equation}

\begin{figure}
\centering
	\includegraphics[scale=1.00]{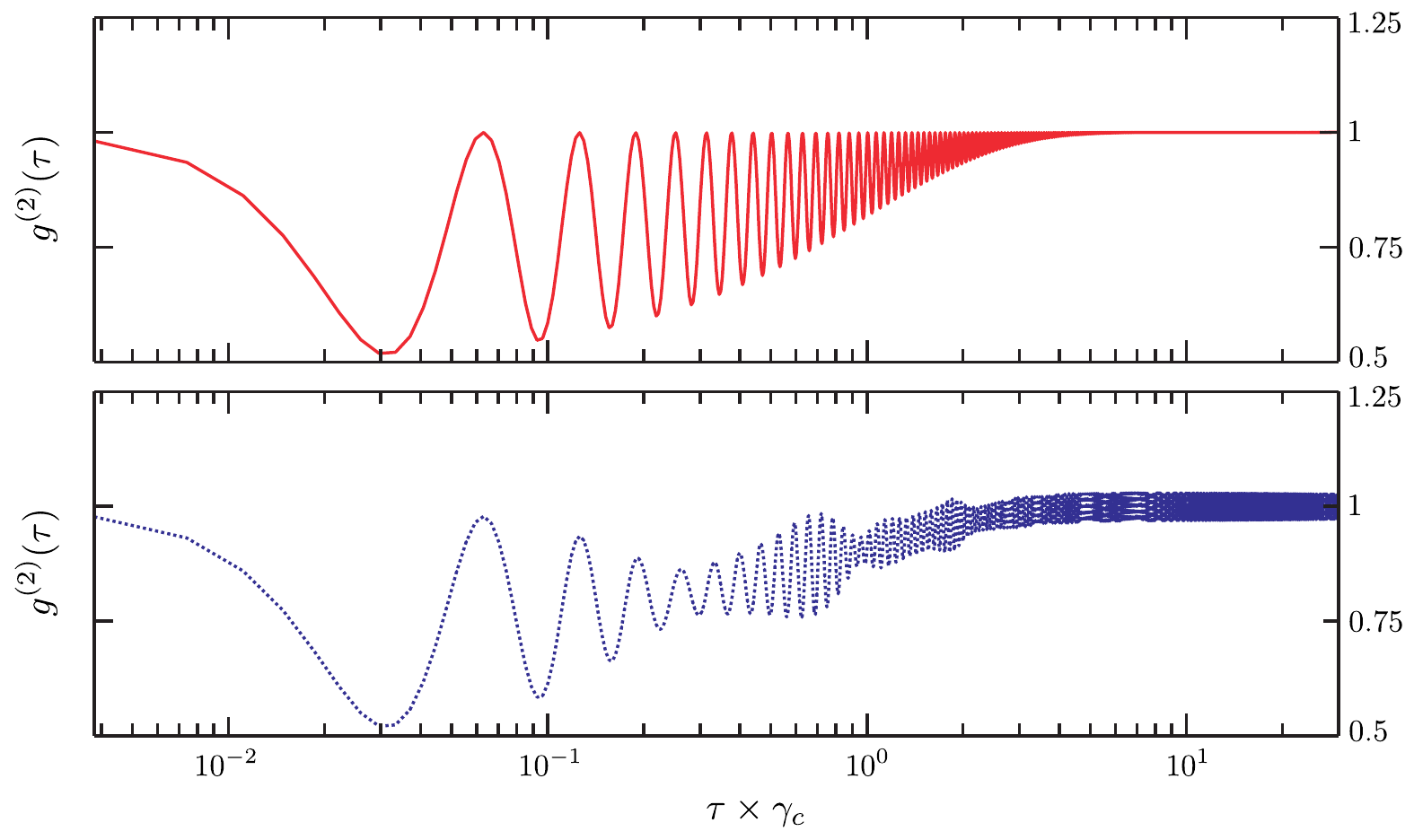}
	\caption{ (Color online) Second-order optical correlation function $g^{(2)}(\tau)$ under incoherent excitation of the atom. The red-solid and blue-dotted lines correspond to $\tilde{g}_{pm} / g_{cm} = 0$ and $1$ respectively. Other parameters:  $ \gamma_{inc}/\gamma_c = 10^ {-3}, ~g_{cm}/\omega_m = 1/10, ~ Q_m=Q_{ac}=10^{4},~ \gamma_c/\gamma_m = 100,~ F_p=0$. 
	The horizontal time axis is plotted in logarithmic scale.}
	\label{fig:g2Tau}
\end{figure}

This quantity has been numerically computed via the quantum regression theorem \cite{Carmichael1999}. \\
Fig. \ref{fig:g2Tau} depicts the behavior of $g^{(2)}(\tau)$
for the case $\tilde{g}_{pm} =0$ (top panel) and $\tilde{g}_{pm} / g_{cm} = 1$ (bottom panel). Given that there are two different
time scales for the cavity photon and mechanical oscillator relaxation times ($\gamma_c/\gamma_m$ = 100), for clarity we have plotted the results using a logarithmic scale for the time axis. For vanishing effective coupling ($\tilde{g}_{pm} =0$), the correlation function $g^{(2)}(\tau)$ oscillates due to the beating of the polariton modes associated to the polariton splitting $2 g_{ac}$ and tends to $1$ on a time scale of the order of $1/\gamma_c$: in this case the mechanical oscillator does not affect the properties as a result of the interference effect.
Notice that for the considered strong coupling parameters, high quality factors and relatively weak incoherent drive,  we find $g^{(2)}(\tau=0) \simeq 1$, in agreement with the analytical formula in Ref. \cite{DelValle},
which has been derived in the limit of weak incoherent drive for a Jaynes-Cummings system consisting of an incoherently pumped two-level emitter coupled to a single cavity mode. On average, $g^{(2)}(\tau)$ stays well below $1$ for $t \gamma_c < 1$.
 
The bottom panel instead reports results for a finite value of $\tilde{g}_{pm}$. For this case, we find that the oscillations become significantly anharmonic due to the existence of the tripartite dressed excitations. Notice that residual oscillations remain up to a scale of the order of $1/\gamma_m$ (they disappear when $t \gg 1/\gamma_m$, not shown). In other words, the effect of the mechanical oscillator affects the long-time behavior of the second-order correlation function.

\begin{figure}
\centering
	\includegraphics[scale=1.00]{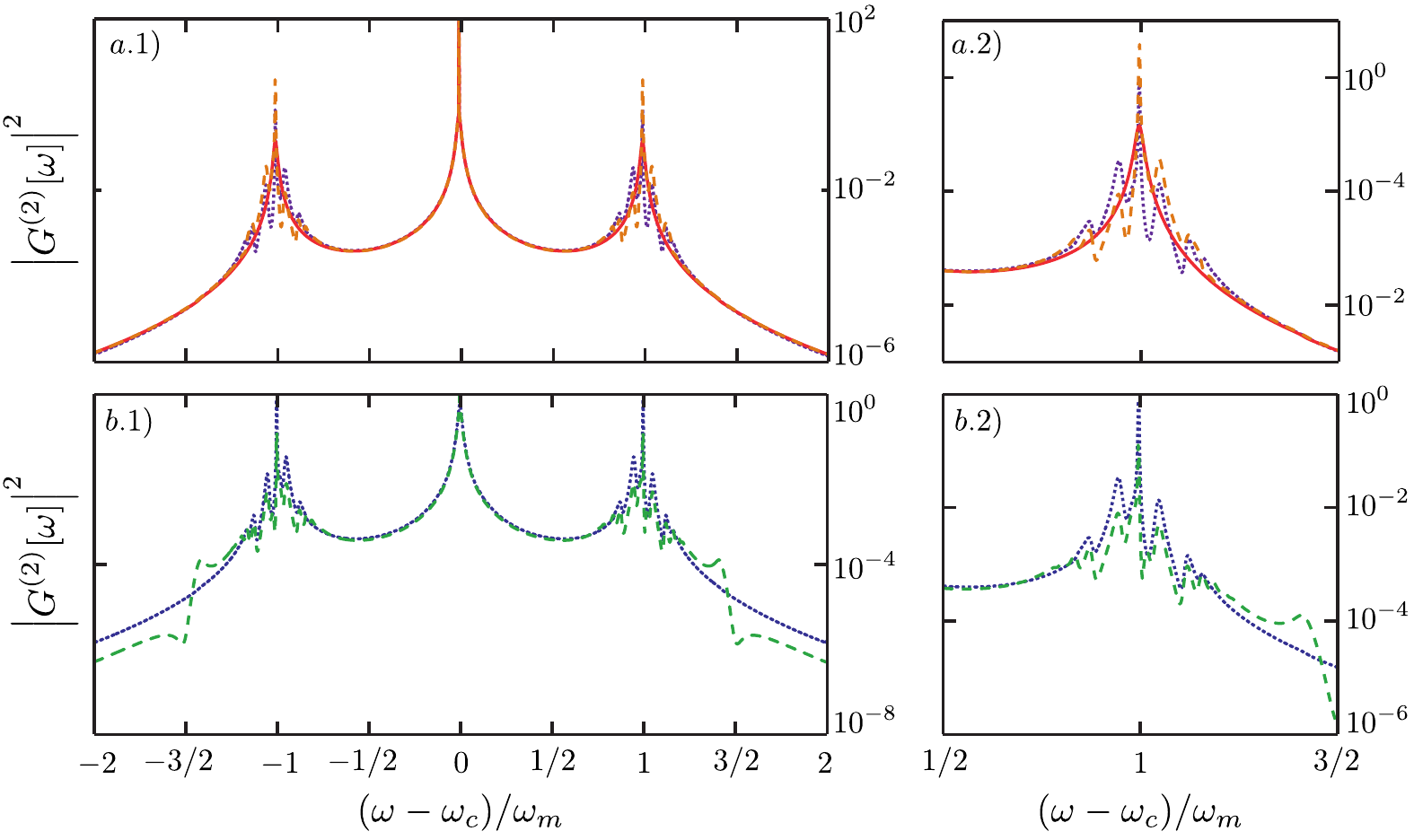}
	\caption{ (Color online) Fourier transform of the second-order optical correlation function $g^{(2)}(\tau)$ of the emission obtained under incoherent excitation of the atom. a) Spectra for different values of the polariton-mechanics effective coupling, $\tilde{g}_{pm} / g_{cm} = -1, 0, 1$ for the blue-dotted, solid-red and dashed-orange lines respectively. Atom incoherent pump rate set to $\gamma_{inc}/\gamma_a=10^{-3}$ a.2) Zoom around $\omega \sim \omega_m = 2 g_{ac}$. b) Spectra for different values of the incoherent pumping rate,  $\gamma_{inc}/\gamma_a=10^{-3},~10^{-1}$ for blue-dotted, green-dashed respectively. The effective polariton coupling has been set to $\tilde{g}_{pm}/g_{cm}=-1$. b.2) Zoom around $\omega \sim \omega_m$. All other parameters have been set to:  $g_{ac}/\omega_m=1/2,~ g_{cm}/\omega_m = 1/10, ~ Q_m=Q_{ac}=10^{4},~ \gamma_c/\gamma_m = 100,~ F_p=0$.}
	\label{fig:opticalEmission}
\end{figure}

Fig. \ref{fig:opticalEmission} presents the spectral function $|G^{(2)}[\omega]|^2$, where  $G^{(2)}[\omega]$ is the Fourier transform of $g^{(2)}(\tau)$  ($G^{(2)}[\omega] = G^{(2)}[-\omega]$ since $g^{(2)}(\tau)$ is real) . In panels  $a.1)$ and $b.1)$,  the central peak at $\omega = 0$ is due to the fact
that $g^{(2)}(\tau) \to 1$ for $\tau \to +\infty$.  The information about the dressed tripartite excitations is contained in the spectral fine structure appearing at $\omega \sim  \omega_m = 2 g_{ac}$ .  Panels $a.2)$ and $b.2)$ display a zoom on the spectral structure centered around $\omega \sim \omega_m = 2 g_{ac} $. The curves in $a.1)$ and $a.2)$ are obtained for an incoherent drive set to $\gamma_{inc}/\gamma_c=10^{-3}$ and for different values of the effective polariton-mechanics coupling. For $\tilde{g}_{pm}=0$ (red-solid line in both insets) there is no fine structure around the peak. When the effective coupling $\tilde{g}_{pm}$ becomes large enough, we observe the appearance of a richer structure associated to the tripartite excitations and discussed extensively in section \ref{sec:effectiveCoupling}. To complete the description, insets $b.1)$ and $b.2)$, present the emission spectra for an effective coupling set to $\tilde{g}_{pm}/g_{cm}=-1$ and for two values of the atom incoherent drive ($\gamma_{inc}/\gamma_a=10^{-3}, 10^{-1}$ corresponding to dotted-blue and dashed green curves respectively). 

\section{Conclusions}
\label{conclusions}
In conclusion, we have studied a fully coupled hybrid system consisting of a two-level atom, an optical cavity and a mechanical resonator. We have determined analytically the dressed excitations of  the corresponding Hamiltonian, when all the three-possible binary interactions are considered. In particular,  we have found how the interplay between atom-mechanics and cavity-mechanics couplings leads to interference patterns in the optomechanical effects on the mechanical resonator. We have also demonstrated the capability of such a system to excite anti-bunched states of mechanical motion under resonant coherent excitation of the cavity. Finally we have discussed the statistical photon emission properties for this kind of hybrid cQED optomechanical systems when excited by an incoherent drive.
Among the future perspectives, an intriguing one is the study of quantum synchronization when considering arrays/lattices of these hybrid cavity optomechanical systems as well as the study of collective phases involving both atom-cavity\cite{light_RMP} and opto-mechanical degrees of freedom.

We acknowledge support from ERC (via the grants GANOMS and CORPHO) and from ANR (via the grant QDOM). 

\end{document}